\documentclass[prd,eqsecnum,leqnum,floats]{revtex4}

\begin{document}

\title{Variation of the light-like particle energy and its critical curve equations}
\author{ W. B. Belayev \\
 \normalsize Center for Relativity and Astrophysics,\\ \normalsize
185 Box, 194358, Sanct-Petersburg, Russia \\e-mail: wbelayev@yandex.ru }

\begin{abstract}
We consider variation of energy of the light-like particle in Riemann
space-time, find lagrangian, canonical momenta and forces. Equations of the
critical curve are obtained by the nonzero energy integral variation in
accordance with principles of the calculus of variations in mechanics. This
method is shown to not lead to violation of conformity of varied curve to the
null path in contradistinction of the interval variation. Though found
equations are differ from standard form of geodesics equations, for the
Schwarzschild space-time their solutions coincide with each other to within
parameter of differentiation.
\end{abstract}

\maketitle


\section{Introduction}

One of postulates of general relativity is claim that in gravity
field in the absence of other forces the word lines of the
material particles and light rays are geodesics.
In differential geometry a geodesic line is defined as a curve, whose tangent vector is parallel propagated along itself \cite{a2}. Differential
equations of geodesic, which is a path of extremal length, can be
found also by the variation method with the aid of the virtual
displacements of coordinates $\mathbf{x}^{i}$ on a small quantity
$\mathbf{\omega}^{i}$. When we add variation to material particle coordinate, the meaning of the time-like interval slow changes, though that leaves it
time-like.

Finding of differential equations of the null geodesic, corresponding to the
light ray motion, by calculus of variations, described in \cite{a1}. In space-time with metrical coefficients $\mathbf{g}_{ij}$ it is considered variation of their first integral
\begin{eqnarray}\label{f1}
h=\mathbf{g}_{ij}\frac{\mathrm{d}\mathbf{x}^{i}}{\mathrm{d}\mu}\frac{\mathrm{d}\mathbf{x}^{j}}{\mathrm{d}\mu}\,,
\end{eqnarray}
where $\mu$ is affine parameter.
Deriving variation
for extremum determination we must admit arbitrary small
displacements of coordinates. The variation of integral of $h$
expanded in multiple Taylor series is written as
\begin{eqnarray}
 \mathrm{\delta}I=\int\limits_{\mu_0}^{\mu_1}\left\{\sum\limits_{n=1}^{\infty}
 \sum\limits_{\beta_{1}+\ldots+\beta_{4}=n}\left[
 \frac{1}{\beta_{1}!\ldots\beta_{4}!}\frac{\mathrm{\partial}^{n}\mathbf{g}_{ij}}{\mathrm{\partial}^{\beta_{1}}\mathbf{x}^{1}
 \ldots\mathrm{\partial}^{\beta_{4}}\mathbf{x}^{4}}
 \frac{\mathrm{d}\mathbf{x}^{i}}{\mathrm{d}\mu}\frac{\mathrm{d}\mathbf{x}^{j}}{\mathrm{d}\mu}(\mathbf{\omega}^{1})^{\beta_1}
 \ldots(\mathbf{\omega}^{4})^{\beta_4}\right]+\mathbf{g}_{ij}\left(2\frac{\mathrm{d}\mathbf{x}^{i}}{\mathrm{d}
\mu}\frac{\mathrm{d}\mathbf{\omega}^{j}}{\mathrm{d}\mu}
 +\frac{\mathrm{d}\mathbf{\omega}^{i}}{\mathrm{d}\mu}\frac{\mathrm{d}\mathbf{\omega}^{j}}{\mathrm{d}\mu}\right)+\right.
\nonumber \\ \left. +\sum\limits_{n=1}^{\infty}
 \sum\limits_{\beta_{1}+\ldots+\beta_{4}=n}\left[
 \frac{1}{\beta_{1}!\ldots\beta_{4}!}\frac{\mathrm{\partial}^{n}\mathbf{g}_{ij}}{\mathrm{\partial}^{\beta_{1}}\mathbf{x}^{1}
 \ldots\mathrm{\partial}^{\beta_{4}}\mathbf{x}^{4}}
\left(2\frac{\mathrm{d}\mathbf{x}^{i}}{\mathrm{d}
\mu}\frac{\mathrm{d}\mathbf{\omega}^{j}}{\mathrm{d}\mu}
 +\frac{\mathrm{d}\mathbf{\omega}^{i}}{\mathrm{d}\mu}\frac{\mathrm{d}\mathbf{\omega}^{j}}{\mathrm{d}\mu}\right)(\mathbf{\omega}^{1})^{\beta_1}
 \ldots(\mathbf{\omega}^{4})^{\beta_4}\right]
\right\}\mathrm{d}\mu,
 \label{f2}
\end{eqnarray}
where $\mu_0, \mu_1$ are meanings of affine parameter in points,
which are linked by found geodesic. With
 finding geodesic equations the
sum of terms containing variations of coordinates and their derivatives by $\mu$
in first power is equated to null, that gives the geodesic equations in form
\begin{eqnarray}
\frac{\mathrm{d}^2 \mathbf{x}^\lambda}{\mathrm{d}\mu ^2}+\Gamma^\lambda_{ij}
\frac{\mathrm{d}\mathbf{x}^i}{\mathrm{d}\mu}\frac{\mathrm{d}\mathbf{x}^j}{\mathrm{d}\mu}=0,
\label{fg}
\end{eqnarray}
where $\Gamma^\lambda_{ij}$ are Christoffel symbols:
\begin{eqnarray}
\Gamma^\lambda_{ij}=\frac{1}{2}\mathbf{g}^{\lambda\gamma}( \mathbf{g}_{i\gamma,j} +\mathbf{g}_{j\gamma,i} - \mathbf{g}_{ij,\gamma}).
\end{eqnarray}
Here a comma denotes partial differentiation.

Other terms of series in (\ref{f2}), containing small quantities
$\mathbf{\omega}^{i}$,
$\mathrm{d}\mathbf{\omega}^{j}/\mathrm{d}\mu$ in more high powers
or their products and being able to have nonzero values, don't
take into account. Thus such method admits violation of condition
$h=0$, which means that with certain coordinates variations the
interval \emph{a prior} becomes time-like or space-like. Since this interval accords with the light ray motion, that leads to the Lorentz-invariance violation in locality, namely, anisotropies.

The possibility of Lorentz symmetry break for the photon in vacuum by effects from the Plank scale is studied in \cite{cpt, anis}. At the contrary, it is shown in \cite{dhs} that a fundamental space-time discreteness need not contradict Lorentz invariance, and causal set's discreteness is in fact locally Lorentz invariant. Experiments \cite{eli} show exceptionally high precision of constancy of light speed confirmed a Lorentz symmetry in locality, and astrophysical tests don't detect isotropic Lorentz violation \cite{anis}.

In the method of calculus of variations in the large \cite{amm} ones are considered as possible paths along the manifold disregarding kind of interval, not as the trajectories of physical particles. This approach exceeds the limits of classical variational principle in mechanics, according as which virtual motions of the system are compared with cinematically possible motions.

Approximating time-like interval conforming in
general relativity to the material particle motion between fixed
points to null leads in physical sense to unlimited increase of
its momentum, and the space-like interval doesn't conform to move of
any object. In this connection it should pay attention on speculation that discreteness at the Planck scale reveals maximum value of momentum for fundamental particles \cite{cam}.

Geodesic line must be extremal \cite{a2}, and the test particle moves along it only in the absence of non-gravity forces. Should photon have some rest mass variations of its path don't give different kinds of intervals, but this assumption doesn't confirm by experiments \cite{a3}. We examine choosing of energy so in order that application of variational principle to its integral for deriving of the isotropic critical curves equations would not lead to considering non-null paths.

\section{Definition of energy and its variation}

The interval in Riemann space-time is written in form
\begin{eqnarray}\label{f21}
\mathrm{d}s^2=\rho^2 \mathbf{g}_{11}\mathrm{d}\mathbf{x}^{12}+
2\rho
\mathbf{g}_{1p}\mathrm{d}\mathbf{x}^{1}\mathrm{d}\mathbf{x}^{p}+
\mathbf{g}_{pq}\mathrm{d}\mathbf{x}^{p}\mathrm{d}\mathbf{x}^{q},
\end{eqnarray}
where $\rho$ is some quantity, whose meaning is assumed to be equal 1. Putting down $\mathbf{x}^1$ as time coordinate, $\mathbf{x}^p \, (p=2,3,4)$ as space coordinates and considering $\rho$ as energy of light-like particle with $\mathrm{d}s=0$ we present it as
\begin{eqnarray}\label{f3}
\rho =\left\lbrace
-\mathbf{g}_{1p}\frac{\mathrm{d}\mathbf{x}^{p}}{\mathrm{d}\mu}+\sigma\left[\left(\mathbf{g}_{1p}\frac{\mathrm{d}\mathbf{x}^{p}}{\mathrm{d}\mu}\right)^{2}-
\mathbf{g}_{11}\mathbf{g}_{pq}\frac{\mathrm{d}\mathbf{x}^{p}}{\mathrm{d}\mu}\frac{\mathrm{d}\mathbf{x}^{q}}{\mathrm{d}\mu}\right]^{1/2}\right\rbrace
\left(\mathbf{g}_{11}\frac{\mathrm{d}\mathbf{x}^{1}}{\mathrm{d}\mu}\right)^{-1}\,,
\end{eqnarray}
where $\sigma$ takes meaning $\pm 1$.

With denotation of the velocity four-vector components as
$\mathbf{v}^{i}=\mathrm{d}\mathbf{x}^{i}/\mathrm{d}\mu$ energy
variation will be
\begin{eqnarray}\label{f31}
\delta \rho= \frac{\partial \rho}{\partial \mathbf{x}^{\lambda}}\delta
\mathbf{x}^{\lambda}+\frac{\partial \rho}{\partial \mathbf{v}^{\lambda}}\delta
\mathbf{v}^{\lambda}
\end{eqnarray}
After substitution

\begin{eqnarray}
\sigma\left[\left(\mathbf{g}_{1p}\frac{\mathrm{d}\mathbf{x}^{p}}{\mathrm{d}\mu}\right)^{2}-
\mathbf{g}_{11}\mathbf{g}_{pq}\frac{\mathrm{d}\mathbf{x}^{p}}{\mathrm{d}\mu}\frac{\mathrm{d}\mathbf{x}^{q}}{\mathrm{d}\mu}\right]^{1/2}=
\mathbf{g}_{1i}\frac{\mathrm{d}\mathbf{x}^{i}}{\mathrm{d}\mu}
\end{eqnarray}
the partial derivatives with respect to coordinates are written as
\begin{eqnarray}\label{f9}
\frac{\mathrm{\partial} \rho }{\mathrm{\partial}
\mathbf{x}^{\lambda}}=\frac{1}{\mathbf{g}_{11}\mathbf{v}^{1}}\left[-\frac{\mathrm{\partial}
\mathbf{g}_{1p}}{\mathrm{\partial}
\mathbf{x}^{\lambda}}\mathbf{v}^{p}+\frac{1}{2\mathbf{v}_{1}}\left(2\frac{\mathrm{\partial}
\mathbf{g}_{1p} }{\mathrm{\partial}
\mathbf{x}^{\lambda}}\mathbf{g}_{1q}-\frac{\mathrm{\partial}
\mathbf{g}_{11} }{\mathrm{\partial}
\mathbf{x}^{\lambda}}\mathbf{g}_{pq}-\frac{\mathrm{\partial}
\mathbf{g}_{pq} }{\mathrm{\partial}
\mathbf{x}^{\lambda}}\mathbf{g}_{11}\right)\mathbf{v}^{p}\mathbf{v}^{q}
\right]- \nonumber \\
-\frac{1}{\mathbf{g}_{11}}\frac{\mathrm{\partial}
\mathbf{g}_{11}}{\mathrm{\partial} \mathbf{x}^{\lambda}}\,.
\end{eqnarray}
This expression is reduced to
\begin{eqnarray}\label{f10}
\frac{\mathrm{\partial} \rho }{\mathrm{\partial}
\mathbf{x}^{\lambda}}=-\frac{1}{2\mathbf{v}_{1}\mathbf{v}^{1}}\frac{\mathrm{\partial}
\mathbf{g}_{ij}}{\mathrm{\partial}
\mathbf{x}^{\lambda}}\mathbf{v}^{i}\mathbf{v}^{j}\,.
\end{eqnarray}
The partial derivatives with respect to components of the velocity four-vector are
\begin{eqnarray}\label{f13}
\frac{\mathrm{\partial} \rho }{\mathrm{\partial}
\mathbf{v}^{\lambda}}=-\frac{\mathbf{v}_{\lambda}}{\mathbf{v}_{1}\mathbf{v}^{1}}\,.
\end{eqnarray}

For the particle, moving in empty space, lagrangian is taken in form
\begin{eqnarray}\label{f130}
L=-\rho,
\end{eqnarray}
and conforms to relation \cite{a4}:
\begin{eqnarray}\label{f131}
\rho=\mathbf{v}^{\lambda}\frac{\mathrm{\partial} L}{\mathrm{\partial}
\mathbf{v}^{\lambda}}-L ,
\end{eqnarray}
which is integral of the motion.
Obtained derivatives give meanings of canonical momenta
\begin{eqnarray}\label{f132}
\mathbf{p}_{\lambda}=\frac{\mathrm{\partial} L}{\mathrm{\partial}
\mathbf{v}^{\lambda}}=\frac{\mathbf{v}_{\lambda}}{\mathbf{v}^{1}\mathbf{v}_{1}}\,
\end{eqnarray}
and forces
\begin{eqnarray}\label{f133}
\mathbf{F}_{\lambda}=\frac{\mathrm{\partial} L }{\mathrm{\partial}
\mathbf{x}^{\lambda}}=\frac{1}{2\mathbf{v}^{1}\mathbf{v}_{1}}\frac{\mathrm{\partial}
\mathbf{g}_{ij}}{\mathrm{\partial}
\mathbf{x}^{\lambda}}\mathbf{v}^{i}\mathbf{v}^{j}\,.
\end{eqnarray}

\section{Equations of isotropic critical curve}

Motion equations are found from
variation of energy integral
\begin{eqnarray}\label{f4}
S=\int\limits_{\mu_0}^{\mu_1}\rho \mathrm{d}\mu,
\end{eqnarray}
where
$\mu_{0}$ and $\mu_{1}$ are fixed points. Energy $\rho$ has non-zero value, its variations leave interval to be light-like, and application of standard variational procedure yields Euler-Lagrange equations

\begin{eqnarray}\label{f8}
\frac{\mathrm{d}}{\mathrm{d}\mu}\frac{\mathrm{\partial}
\rho }{\mathrm{\partial} \dot{\mathbf{x}}^{\lambda}}-\frac{\mathrm{\partial} \rho }{\mathrm{\partial}
\mathbf{x}^{\lambda}}=0\,.
\end{eqnarray}

Critical curve equations are obtained by substitution of partial derivatives (\ref{f10}) and (\ref{f13}) in these equations. For the time coordinate we have
\begin{eqnarray}\label{f12}
\frac{\mathrm{d}\mathbf{v}^{1}}{\mathrm{d}\mu}+\frac{\mathbf{v}^{1}}{2\mathbf{v}_{1}}\frac{\mathrm{\partial}
\mathbf{g}_{ij}}{\mathrm{\partial}
\mathbf{x}^{1}}\mathbf{v}^{i}\mathbf{v}^{j}=0\,.
\end{eqnarray}
For finding of other three equations of motion the second term of (\ref{f8}) is presented in form
\begin{eqnarray}
\frac{\mathrm{d}}{\mathrm{d}\mu}\frac{\mathrm{\partial}
c}{\mathrm{\partial}
\mathbf{v}^{\lambda}}=\frac{1}{\mathbf{v}^{1}(\mathbf{v}_{1})^{2}}\left[(\mathbf{g}_{1p}\mathbf{v}_{\lambda}-\mathbf{g}_{p\lambda}
\mathbf{v}_{1})
\frac{\mathrm{d}\mathbf{v}^{p}}{\mathrm{d}\mu}-\left(\frac{\mathrm{\partial}
\mathbf{g}_{i\lambda}}{\mathrm{\partial}
\mathbf{x}^{j}}\mathbf{v}_1-\frac{\mathrm{\partial}
\mathbf{g}_{1i}}{\mathrm{\partial}
\mathbf{x}^{j}}\mathbf{v}_{\lambda}\right)\mathbf{v}^i \mathbf{v}^j
\right]+
 \nonumber \\
 +\frac{\mathbf{g}_{11}\mathbf{v}^{1}\mathbf{v}_{\lambda}+\mathbf{g}_{p\lambda} \mathbf{v}^{p}\mathbf{v}_{1}}{(\mathbf{v}^{1})^{2}\mathbf{v}_{1}^{2}}\frac{\mathrm{d}\mathbf{v}^{1}}{\mathrm{d}\mu}\,.
\label{f14}
\end{eqnarray}
Replacement of derivative $\mathrm{d}\mathbf{v}^{1}/\mathrm{d}\mu$
here on its expression, obtained from (\ref{f12}), and
substitution found terms in Euler-Lagrange equations gives
\begin{eqnarray}
(\mathbf{g}_{p\lambda}
\mathbf{v}_{1}-\mathbf{g}_{1p}\mathbf{v}_{\lambda})\frac{d\mathbf{v}^{p}}{d\mu}+\left[\frac{1}{2\mathbf{v}_{1}}\frac{\mathrm{\partial}
\mathbf{g}_{ij}}{\mathrm{\partial}
\mathbf{x}^{1}}(\mathbf{g}_{11}\mathbf{v}^{1}\mathbf{v}_{\lambda}+\mathbf{g}_{p\lambda}\mathbf{v}^{p}\mathbf{v}_{1})-\frac{1}{2}\frac{\mathrm{\partial}
\mathbf{g}_{ij}}{\mathrm{\partial}
\mathbf{x}^{\lambda}}\mathbf{v}_{1} +\frac{\mathrm{\partial}
\mathbf{g}_{i\lambda}}{\mathrm{\partial}
\mathbf{x}^{j}}\mathbf{v}_1- \frac{\mathrm{\partial}
\mathbf{g}_{1i}}{\mathrm{\partial}
\mathbf{x}^{j}}\mathbf{v}_{\lambda}\right]\mathbf{v}^i \mathbf{v}^j=0\,.
\label{f15}
\end{eqnarray}
These equations contain accelerations corresponded to the space coordinates and coupled with (\ref{f12}) describe motion of the test light-like particle along critical curve. They don't coincide to usual form (\ref{fg}) of the null geodesics equations.

\section{Isotropic curves in Schwarzschild space-time}

Central gravity field of a point body is described by the Schwarzschild metric. This is solution of Einstein's field equations for space-time, which is empty except central point. We chose units so, that a light velocity constant $c$ is rendered to $1$. At spherical coordinates $\mathbf{x}^i=(t,r,\theta,\varphi)$ the line element is
\begin{eqnarray}\label{f16}
\mathrm{d}s^{2}=\left(1-\frac{\alpha}{r}\right)\mathrm{d}t^{2}-\left(1-\frac{\alpha}{r}\right)^{-1}\mathrm{d}r^{2}-
r^{2}(\mathrm{d}\theta^{2}+{\sin}^{2}\theta
\mathrm{d}\varphi^{2}),
\end{eqnarray}
where $\alpha$ is constant.
Motion equations (\ref{f12}) and (\ref{f15}) lead to
\begin{eqnarray}\label{f17}
\frac{\mathrm{d}^{2}t}{\mathrm{d}\mu^2}=0
\end{eqnarray}
and
\begin{eqnarray}\label{f18}
\frac{\mathrm{d}^{2}r}{\mathrm{d}\mu^2}+\frac{\alpha}{2r^2}\left(1-
\frac{\alpha}{r}\right)\left(\frac{\mathrm{d}t}{\mathrm{d}\mu}\right)-
\frac{3\alpha}{2r(r-\alpha)}\left(\frac{\mathrm{d}r}{\mathrm{d}\mu}\right)^2-
 \nonumber \\
-(r-\alpha)\left[\left(\frac{\mathrm{d}\theta}{\mathrm{d}\mu}\right)^2+
{\sin}^{2}\theta\left(\frac{\mathrm{d}\varphi}{\mathrm{d}\mu}\right)^2\right]=0\,,
\end{eqnarray}
\begin{eqnarray}\label{f19}
\frac{\mathrm{d}^{2}\theta}{\mathrm{d}\mu^2}+\frac{2r-3\alpha}{r(r-\alpha)}\frac{\mathrm{d}r}{\mathrm{d}\mu}\frac{\mathrm{d}\theta}{\mathrm{d}\mu}+
\frac{1}{2}{\sin}
2\theta\left(\frac{\mathrm{d}\varphi}{\mathrm{d}\mu}\right)^2=0\,,
\end{eqnarray}
\begin{eqnarray}\label{f20}
\frac{\mathrm{d}^{2}\varphi}{\mathrm{d}\mu^2}+\frac{2r-3\alpha}{r(r-\alpha)}\frac{\mathrm{d}r}{\mathrm{d}\mu}\frac{\mathrm{d}\varphi}{\mathrm{d}\mu}+
2\cot\theta\frac{\mathrm{d}\theta}{\mathrm{d}\mu}\frac{\mathrm{d}\varphi}{d\mu}=0\,.
\end{eqnarray}
Metric (\ref{f16}) for the isotropic curve yields
\begin{eqnarray}\label{f21}
\left(1-\frac{\alpha}{r}\right)\left(\frac{\mathrm{d}t}{\mathrm{d}\mu}\right)^{2}
-\left(1-\frac{\alpha}{r}\right)^{-1}\left(\frac{\mathrm{d}r}{\mathrm{d}\mu}\right)^{2}-
r^2\left[\left(\frac{\mathrm{d}\theta}{\mathrm{d}\mu}\right)^2+
{\sin}^{2}\theta\left(\frac{\mathrm{d}\varphi}{\mathrm{d}\mu}\right)^2\right]=0\,.
\end{eqnarray}

Equations (\ref{f17}) and (\ref{f20}) are solvable by quadrature:
\begin{eqnarray}\label{f22}
\frac{\mathrm{d}t}{\mathrm{d}\mu}=A\,,
\end{eqnarray}
\begin{eqnarray}\label{f23}
\frac{\mathrm{d}\varphi}{\mathrm{d}\mu}=\frac{h}{r^{2}{\sin}^{2}\theta}\left(1-\frac{\alpha}{r}\right)
\,,
\end{eqnarray}
where $A,\,h$ are constants. Substituting
obtained solutions in equation (\ref{f21}) with $\theta=\pi/2$ we find
\begin{eqnarray}\label{f24}
\frac{\mathrm{d}r}{\mathrm{d}\mu}=\pm\left[\left(1-\frac{\alpha}{r}\right)^2
A^2-\left(\frac{h}{r}\right)^2\left(1-\frac{\alpha}{r}\right)^3\right]^{1/2}\,.
\end{eqnarray}
Though found equations (\ref{f12}) and (\ref{f15}) for the
Schwarzschild space-time are differ from standard form of null
geodesic equations, their solutions coincide with each other to within parameter of differentiation
\begin{eqnarray}\label{f25}
\mathrm{d}\mu=\mathrm{d}\mu_s\left(1-\frac{\alpha}{r}\right)^{-1},
\end{eqnarray}
where $\mu$ corresponds with standard solution \cite{a1}.

If we choose $\mu=t$ it follows from Eq. (\ref{f22}) that $A=1$. Canonical momenta (\ref{f132}) and forces (\ref{f133}) under considering constraint on $\theta$ are

\begin{eqnarray}\label{f251}
\mathbf{p}_{1}=1,\,\mathbf{p}_{2}=\mp\frac{1}{\left(1-\frac{\alpha}{r}\right)}
\sqrt{1-\frac{h^{2}}{r^{2}}\left(1-\frac{\alpha}{r}\right)},\,\mathbf{p}_{3}=0,\,
\mathbf{p}_{4}=-h;
\end{eqnarray}

\begin{eqnarray}\label{f252}
\mathbf{F}_{1}=\mathbf{F}_{3}=\mathbf{F}_{4}=0,\,
\mathbf{F}_{2}=\frac{\alpha}{r^{2}\left(1-\frac{\alpha}{r}\right)}-\frac{h^{2}}{r^{3}}+
\frac{\alpha h^2}{2r^{4}}.
\end{eqnarray}
A nonzero component of the contrvariant vector of canonical force is
\begin{eqnarray}\label{f253}
\mathbf{F}^{2}=-\frac{\alpha}{r^{2}}+\frac{h^{2}}{r^{3}}\left(1-\frac{\alpha}{r}\right)\left(1-\frac{\alpha}{2r}\right).
\end{eqnarray}
In so far as Newtonian limit of gravity theory with gravitational constant $G$ and mass $M$ requires $\alpha=2GM$, the first term of $\mathbf{F}^{2}$ yields twice Newton gravity force. That conforms to light deflection in central gravity field \cite{a34}, which is twice meaning being given by Newton gravity theory.

\section{Conclusion}

Presented form of energy allows applying of method of lagrangian for analysis of light-like particle motion. Virtual displacements of coordinates retain conformity of the trajectory of the light-like particle to the null path in Riemann space-time, not leaded to Lorentz-invariance violation. Obtained extremal isotropic curve equations for the Schwarzschild metric give the same solution, which follows from standard null geodesics equations to within appropriate parameter. At that found equations as whole are different from one, which are produced by variational method being applied to the path.

\end{document}